\documentclass[11pt]{revtex4}
\usepackage{amsfonts}
\usepackage{amssymb,epsf}
\usepackage{latexsym}
\begin{document}
\vspace*{1cm}
\begin{flushright}
hep-th/0701198
\end{flushright}
\title{Thermodynamical Properties of Apparent Horizon \\
in Warped DGP Braneworld}
\author{Ahmad Sheykhi $^{1,2}$\footnote{Email:
asheykhi@mail.uk.ac.ir} and Bin Wang $^{1}$\footnote{Email:
wangb@fudan.edu.cn}}
\address{$^1$  Department of Physics, Fudan University, Shanghai 200433, China\\
         $^2$  Physics Department and Biruni Observatory, Shiraz University, Shiraz 71454, Iran}

\author{Rong-Gen Cai\footnote{Email: cairg@itp.ac.cn}} \affiliation{Institute of Theoretical Physics,
Chinese Academic of Sciences, \\  P.O. Box 2735, Beijing 100080,
China}

\begin{abstract}
\vspace*{1cm} \centerline{\bf Abstract}
 In this paper we first obtain Friedmann equations
for the $(n-1)$-dimensional brane embedded in the
$(n+1)$-dimensional bulk, with intrinsic curvature term of the
brane included in the action (DGP model). Then, we show that one
can always rewrite the Friedmann equations in the form of the
first law of thermodynamics, $dE=TdS+WdV$, at apparent horizon on
the brane, regardless of whether there is the intrinsic curvature
term on the brane or a cosmological constant in the bulk. Using
the first law, we extract the entropy expression of the apparent
horizon on the brane. We also show that in the case without the
intrinsic curvature term, the entropy expressions are the same by
using the apparent horizon on the brane and by using the bulk
geometry.  When the intrinsic curvature appears, the entropy of
apparent horizon on the brane has two parts,
 one part follows the $n$-dimensional area formula on the brane, and
the other part is the same as the entropy in the case without the
intrinsic curvature term.  As an interesting result, in the warped
DGP model, the entropy expression in the bulk and on the brane are
not the same. This is reasonable, since in this model gravity on
the brane has two parts, one induced from the $(n+1)$-dimensional
bulk gravity and the other due to the intrinsic curvature term on
the brane.
\end{abstract}

 \maketitle
 \newpage
\section{Introduction}\label{Intr}

The profound connection between the black hole physics and
thermodynamics revealed in the 1970s' inspired deep thinking on the
relation between gravity and thermodynamics in general. A pioneer
work on this respect was done by Jacobson a decade ago who showed
that the gravitational Einstein equation can be derived from the
relation between the horizon area and entropy, together with the
Clausius relation $\delta Q=T\delta S$ \cite{Jac}. The study on the
relation between gravity and thermodynamics has been extended beyond
the Einstein gravity. For the so called $f(R)$ gravity, Eling
\textit{et al.} \cite{Elin} recently argued that the corresponding
field equation describing gravity can be derived from thermodynamics
by using the procedure in \cite{Jac}, but a treatment with
nonequilibrium thermodynamics of spacetime is needed. In the
framework of Gauss-Bonnet gravity, the scalar-tensor gravity and
more general Lovelock gravity, this problem has also been studied in
\cite{Cai1}\cite{Pad} respectively.

It is of great interest to generalize the study to the cosmological
context. Attempts to disclose the connection between Einstein
gravity and thermodynamics in the cosmological situation have been
carried out in \cite{Cai2,Cai3,CaiKim,Fro,Dan,verlinde}. It has been
shown that the differential form of the Friedmann equation of the
FRW universe at the apparent horizon can be rewritten in the form of
the first law of thermodynamics~\cite{Cai2}. Besides the usual FRW
universe, it is also of interest to explore the deep connection
between gravity and thermodynamics in the brane world
cosmology\cite{Cai4}.

In recent years, there are a lot of interest in the brane world
scenario, based on the assumption that all matters in standard
model of particle physics are confined on a surface (brane)
embedded in a higher dimensional spacetime (bulk), while the
gravitational field, in contrast, is usually considered to live in
the whole spacetime. There are two main pictures in the brane
world scenario. In the first picture which we refer as the
Randall-Sundrum II model (RS II), a positive tension 3-brane
embedded in  an 5-dimensional AdS bulk and the cross over between
4D and 5D gravity is set by the AdS radius \cite{RS}. In this
case, the extra dimension has a finite size. In another picture
which is based on the work of  Dvali, Gabadadze, Porrati (DGP
model)\cite{DGP,DG}, a 3-brane is embedded in a spacetime with an
infinite-size extra dimension, with the hope that this picture
could shed new light on the standing problem of the cosmological
constant as well as on supersymmetry breaking \cite{DGP,Wit}. The
recovery of the usual gravitational laws in this picture is
obtained by adding to the action of the brane an Einstein-Hilbert
term computed with the brane intrinsic curvature. The presence of
such a term in the action is generically induced by quantum
corrections coming from the bulk gravity and its coupling with
matter living on the brane and should be included in a large class
of theories for self-consistency \cite{DG, Adler}.

It is worth applying the method developed in \cite{Cai2,Cai3,Cai4}
to investigate the connection between gravity and thermodynamical
properties of the apparent horizon of the universe in the framework
of brane world scenarios. Gravity on the brane does not obey
Einstein theory, thus the usual area formula for the black hole
entropy does not hold on the brane. In addition, exact analytic
black hole solutions on the brane have not been found until now, so
that the relation between the braneworld black hole horizon entropy
and it's geometry is  not known. It is expected that the connection
between gravity and thermodynamics in the braneworld can shed some
lights on understanding these problems. This is our main motivation
to explore the thermodynamical properties of the Friedmann equation
in the warped braneworld. Exact Friedmann equations on the brane for
the RS II model have been derived by many authors (see e.g
\cite{BDEL}). In this paper we will show that the differential form
of the Friedmann equations on the $(n-1)$-dimensional brane embedded
in an $(n+1)$-dimensional Minkowski or AdS  spacetime can be written
directly in the form of the first law of thermodynamics,
$dE=TdS+WdV$, on the apparent horizon. Using the first law, we will
extract the entropy expression of the apparent horizon on the brane,
which coincides with the result in \cite{Cai4} obtained by using the
method of unified first law and the Clausius relation. When the
intrinsic curvature appears on the brane, we will show that the
entropy of apparent horizon on the brane is a sum of two terms, one
is the area formula on the brane and the other is the entropy
expression in the case without the intrinsic curvature term.

The outline of the paper is as follows. In Sec.~\ref{GF}, we
generalize Friedmann equations for the $(n-1)$-dimensional brane
embedded in an $(n+1)$-dimensional bulk, with intrinsic curvature
term of the brane included in the action. In Sec.~\ref{RS}, we study
thermodynamical behavior of Friedmann equation in RS II braneworld
scenario and extract the entropy of apparent horizon from the first
law. In Sec.~\ref{DGP}, we extend our method for the warped DGP
brane world scenario and find out the entropy associated with
apparent horizon on the brane. The last section is devoted to
summary and conclusions.

\section{General Formalism }\label{GF}

We consider an $(n-1)$-dimensional brane embedded in an
$(n+1)$-dimensional spacetime with an intrinsic curvature term
included in the brane action
\begin{eqnarray}\label{Act}
I_{G} &=&-\frac{1}{2\kappa_{n+1}^2 } \int {d^{n+1}x
\sqrt{-\widetilde{g}}\widetilde{R}} +\int
{d^{n+1}x\sqrt{-\widetilde{g}}\mathcal {L}_m}
-\frac{1}{2\kappa_{n}^2 } \int {d^{n}x \sqrt{-g}{R}}.
\end{eqnarray}
The first term in (\ref{Act}) corresponds to the Einstein-Hilbert
action in the $(n+1)$-dimensional bulk, where
 $\widetilde{g}_{AB}$ is the bulk metric and $\widetilde{R}$ is
 the $(n+1)$-dimensional scalar curvature. Similarly, the
last term is the Einstein-Hilbert action for the induced metric
$g_{\mu\nu}$ on the brane with scalar curvature $R$.  The second
term in (\ref{Act}) corresponds to the matter content. Aside from
the bulk matter, we have included the contribution of the
brane-localized matter, which can be rewritten as
\begin{equation}
\int {d^{n}x\sqrt{-g}( l_{m} -2 \lambda)}
\end{equation}
where $l_m$ is the lagrangian density of the brane matter fields,
and $\lambda$ is the brane tension (or cosmological constant). We
emphasize here that this tension is not related to the presence of
the intrinsic curvature term, and can in principle be tuned to be
zero \cite{DGP,DG}. Hereafter we assume that the brane
cosmological constant is zero (if it does not vanish, one can
absorb it in the stress-energy tensor of perfect fluid on the
brane) and redefine
\begin{equation}\label{rela}
 \kappa_{n+1}^2=8\pi
 G_{n+1}\, ,\quad \kappa_{n}^2=8\pi
 G_{n}\, ,\quad
 \Lambda_{n+1}=-\frac{n(n-1)}{2\kappa_{n+1}^2\ell^2},
\end{equation}
where $\Lambda_{n+1}$ is the $(n+1)$-dimensional bulk cosmological
constant. We will consider $(n+1)$-dimensional spacetime metric of
the form
\begin{eqnarray}
ds^2 = \widetilde {g}_{AB}dx^A dx^B &=& g_{\mu\nu}dx^{\mu}
dx^{\nu} + dy ^2
\end{eqnarray}
where $y$ is the coordinate of the bulk. We assume the brane is
located at $y=0$ and the bulk has $\mathbb{Z}_2$ symmetry. Since
we are interested in cosmological solutions, we take the metric in
the form
\begin{eqnarray}
ds^2 = -N^2(t,y) dt^2+ A^2(t,y)\gamma_{ij}dx^i dx^j+ dy^2
\end{eqnarray}
where $\gamma _{ij}$ is a maximally symmetric $(n-1)$-dimensional
metric for the surface ($t$=const., $y$=const.), whose spatial
curvature is parameterized by k = -1, 0, 1. On every hypersurface
($y$=const), we have the metric of a FRW cosmological model. The
metric coefficients $A$ and $N$ are chosen so that, $N(t,0)=1$,
$A(t,0)=a(t)$ and $t$ is cosmic time on the brane. The
$(n+1)$-dimensional Einstein equations take the form
\begin{equation}
\widetilde{G}_{AB}\equiv \widetilde{R}_{AB}-\frac{1}{2}\widetilde
{R} \widetilde{g}_{AB}=\kappa_{n+1}^2 \widetilde{S}_{AB} .
\label{Ein}
\end{equation}
where $\widetilde{R}_{AB}$ is the $(n+1)$-dimensional Ricci tensor,
and the tensor $\widetilde{S}_{AB}$ is the sum of the
energy-momentum tensor $\widetilde{T}_{AB}$ of matter and the
contribution coming from the scalar curvature of the brane. We
denote this latter contribution by $\widetilde{U}_{AB}$. We have
\begin{equation}
\widetilde{S}_{AB}= \widetilde{T}_{AB}+\widetilde{U}_{AB}
\end{equation}
The energy-momentum tensor can be further decomposed into two
parts
\begin{equation}
\widetilde{T}_{AB}= -\Lambda_{n+1}\widetilde{g}_{AB}+{T}_{AB},
\end{equation}
where ${T}_{AB}$, is the matter content on the brane $(y=0)$,
which we assume in the form of a perfect fluid for homogenous and
isotropic universe on the brane
\begin{equation}
T_{AB}= \delta^{\mu}_{A}\delta^{\nu}_{B}t_{\mu \nu}\delta (y)\,
,\quad t_{\mu \nu}=(\rho+p)u_{\mu}u_{\nu}+pg_{\mu \nu},
\end{equation}
where $u^{\mu}$, $\rho$ and $p$ being the perfect fluid velocity
($u^{\mu}u_{\nu}=-1$), energy density and pressure respectively.
The possible contribution of a nonzero brane tension $\lambda$
will be assumed to be included in $\rho$ and $p$. The assumption
that $\widetilde {T}_{0y} = 0$, which physically means that there
is no energy flow between the brane and the bulk, implies that
$\widetilde{G}_{0y} = 0$ vanishes. The nonvanishing components of
$\widetilde{U}_{AB}$ are
\begin{equation}
\widetilde{U}_{00}=-\frac{\delta(y)}{2\kappa_{n}^2}(n-1)(n-2)\left(\frac{\dot{A}^2}{A^2}+k\frac{N^2}{A^2}\right),
\end{equation}
\begin{equation}
\widetilde{U}_{ij}=-\frac{\delta(y)}{2\kappa_{n}^2}(n-2)(n-3)\gamma_{ij}\left[
\frac{A^2}{N^2}\left(-\frac{\dot{A}^2}{A^2}+\frac{2}{n-3}\frac{\dot{N}\dot{A}}{N
A}-\frac{2}{n-3}\frac{\ddot{A}}{A}\right)-k \right],
\end{equation}
where dot denotes derivative with respect to $t$. Our aim here is to
obtain the Friedmann equation governing the cosmological evolution
on the brane. Following \cite{BDEL} we find that a set of functions
$A(t,y)$ and $N(t,y)$  satisfying in equation
\begin{equation}\label{Field}
\left(\frac{\dot{A}}{NA}\right)^2+\frac{k}{A^2}
-\frac{2\kappa_{n+1}^2\Lambda_{n+1}}{n(n-1)}-\left(\frac{A'}{A}\right)^2-\frac{\mathcal{C}}{A^n}=0,
\end{equation}
together with $\widetilde{G}_{0y} = 0$ which leads
$\dot{A}/N=\alpha(t)$, will be solutions of the field equations in
the bulk. In Eq. (\ref{Field}) prime denotes derivative with respect
to $y$ and $\mathcal{C}$ is an integration constant which is related
to the $(n+1)$-dimensional bulk Weyl tensor~\cite{Shiro,Muk} and
will be zero in the cases of Minkowski and AdS bulk. To find the
Friedmann equation on the brane we need to study the junction
condition on the brane. The metric is required to be continuous
across the brane. However its derivative with respect to $y$ can be
discontinuous at $y = 0$. This will entail the existence of a Dirac
delta function in the second derivative of the metric with respect
to $y$ (see \cite{BDL} for details). Integrating the $(00)$ and
$(ij)$ components of the field equations (\ref{Ein}) across the
brane and imposing $\mathbb{Z}_2$ symmetry, the Junction conditions
are shown as follows
\begin{equation}\label{Jc1}
\frac{2{a}^{\prime}}{a}=-\frac{\kappa_{n+1}^2
}{n-1}\rho+\frac{\kappa_{n+1}^2(n-2)}{2\kappa_{n}^2}\left(\frac{\dot{a}^2}{a^2}+\frac{k}{a^2}\right),
\end{equation}
\begin{equation}\label{Jc2}
2N^{\prime}_{+}=\kappa_{n+1}^2\left(p
+\frac{n-2}{n-1}\rho\right)-\frac{\kappa_{n+1}^2(n-2)}{2\kappa_{n}^2}\left(
\frac{\dot{a}^2}{a^2}+\frac{k}{a^2}-2\frac{\ddot{a}}{a}\right),
\end{equation}
where $2{N^{\prime}_{+}}=-2{N^{\prime}_{-}}$ is the discontinuity
of the first derivative. One may note that in the particular case
$n=4$ these junction conditions reduce to those obtained in
\cite{Def} for $3$-brane embedded in a $5$-dimensional bulk
including intrinsic curvature term on the brane (DGP model). On
the other hand taking the limit $\kappa_{n}\to \infty$ we obtain
the junction conditions on the $(n-1)$-brane embedded in an
$(n+1)$-dimensional bulk in RS II brane scenario. Using
(\ref{Jc1}) in Eq. (\ref{Field}) on the brane $(y=0)$ we have the
 generalized Friedmann equation
\begin{equation}\label{GFri}
\epsilon
\sqrt{H^2+\frac{k}{a^2}-\frac{2\kappa_{n+1}^2\Lambda_{n+1}}{n(n-1)}-\frac{\mathcal{C}}{a^n}}
=-\frac{\kappa_{n+1}^2}{4\kappa_{n}^2}(n-2)(H^2+\frac{k}{a^2})+\frac{\kappa_{n+1}^2}{2(n-1)}\rho,
\end{equation}
where $H=\dot{a}/a$ is the Hubble parameter on the brane and
$\epsilon=\pm1$. For later convenience we choose $\epsilon=1$.
Inserting boundary conditions (\ref{Jc1}) and (\ref{Jc2}) into the
equation $\widetilde{G}_{0y} = 0$, we get the continuity equation
for the perfect fluid confined on the brane
\begin{equation}
\label{Cont}
 \dot{\rho}+(n-1)H(\rho+p)=0.
\end{equation}
Eqs. (\ref{GFri}) and (\ref{Cont}), together with the equation of
state $p=p(\rho)$, describe completely  the cosmological dynamics
on the brane. We will use this equation and the Friedmann equation
to obtain the  first law of thermodynamics at the apparent horizon
on the brane both in the RS II model and the DGP model.

To have further understanding about the nature of apparent horizon
we rewrite more explicitly, the metric of homogenous and isotropic
FRW universe on the brane in the form
\begin{equation}
ds^2={h}_{\mu \nu}dx^{\mu} dx^{\nu}+\tilde{r}^2d\Omega_{n-2}^2,
\end{equation}
where $\tilde{r}=a(t)r$, $x^0=t, x^1=r$, the two dimensional
metric $h_{\mu \nu}$=diag $(-1, a^2/(1-kr^2))$ and $d\Omega_{n-2}$
is the metric of $(n-2)$-dimensional unit sphere. Then, the
dynamical apparent horizon, a marginally trapped surface with
vanishing expansion, is determined by the relation $h^{\mu
\nu}\partial_{\mu}\tilde {r}\partial_{\nu}\tilde {r}=0$, which
implies that the vector $\nabla \tilde {r}$ is null on the
apparent horizon surface. The apparent horizon has been argued to
be a causal horizon for a dynamical spacetime and is associated
with gravitational entropy and surface gravity \cite{Hay2,Bak}.
The explicit evaluation of the apparent horizon for the FRW
universe gives the apparent horizon radius
\begin{equation}
\label{radius}
 \tilde{r}_A=\frac{1}{\sqrt{H^2+k/a^2}}.
\end{equation}
The associated temperature on the apparent horizon can be defined
as $T = \kappa/2\pi$, where $\kappa$ is the surface gravity
\begin{equation}
\label{surgra}\label{kappa}
 \kappa =\frac{1}{\sqrt{-h}}\partial_{a}\left(\sqrt{-h}h^{ab}\partial_{ab}\tilde
 {r}\right),
\end{equation}
Then one can easily show that the surface gravity at the apparent
horizon of FRW universe can be written as
\begin{equation}\label{surgrav}
\kappa=-\frac{1}{\tilde r_A}\left(1-\frac{\dot {\tilde
r}_A}{2H\tilde r_A}\right).
\end{equation}
In the remain parts of this paper we will apply the result of this
section to investigate the thermodynamical properties of apparent
horizon in the braneworld scenario. We will show that the first
law of thermodynamics on the apparent horizon can be extracted
directly from the Friedmann equations for both AdS and Minkowski
bulk and therefore we can extract an entropy relation on the
brane.

\section{Thermodynamics of Apparent Horizon in RS II model}\label{RS}

Let us begin by RS II model in which no intrinsic curvature term
on the brane contributes in the action. Taking the limit
$\kappa_{n}\to \infty$, while keeping $\kappa_{n+1}$ finite, the
equation (\ref{GFri}) reduces to the Friedmann equation in the RS
II brane world model
\begin{equation}\label{RSFri}
H^2+\frac{k}{a^2}-\frac{2\kappa_{n+1}^2\Lambda_{n+1}}{n(n-1)}-\frac{\mathcal{C}}{a^n}
=\frac{\kappa_{n+1}^4}{4(n-1)^2}\rho^2.
\end{equation}
If one invokes the standard assumption that the energy density on
the brane can be separated into two contributions, the ordinary
matter component,$\rho_{b}$, and the brane tension, $\lambda > 0$,
such that $\rho= \rho_{b}+\lambda$, (after fine tuning between the
brane tension and the bulk cosmological constant), then one can
recover the Friedmann equation presented in \cite{Cai4} for
$n$-dimensional RS braneworld Scenario. We recall that one can
interpret the constant $\mathcal{C}$ coming from the
$n+1$-dimensional bulk Weyl tensor. Since we are interested here in
flat (Minkowskian) and conformally flat (AdS) bulk spacetime, the
bulk Weyl tensor will vanish, so we set $\mathcal{C}=0$ in the
following discussions.

\subsection{Brane embedded in Minkowski bulk}\label{minRS}
We begin by the simplest case, namely Minkowski bulk, in which
$\Lambda_{n+1}=0$, so we can rewrite the Friedmann equation
(\ref{RSFri}) in the simple form
\begin{equation}
\label{minRS1}
 H^2+\frac{k}{a^2}=\frac{\kappa_{n+1}^4 }{4 (n-1)^2}\rho^2.
 \end{equation}
In terms of the apparent horizon radius, we can rewrite the
Friedmann equation (\ref{minRS1}) on the brane as
\begin{equation}
\label{minRS2}
 \frac{1}{\tilde {r}_{A}}=\frac{4\pi G_{n+1}}{n-1} \rho,
 \end{equation}
where we have used Eq. (\ref{rela}). Taking differential form of
equation (\ref{minRS2}) and using the continuity equation
(\ref{Cont}), one can get the differential form of the Friedmann
equation on the brane
\begin{equation} \label{minRS3}
\frac{1}{4\pi G_{n+1}} \frac{d\tilde {r}_{A}}{\tilde
{r}_{A}^2}=H(\rho+p)dt.
\end{equation}
Multiplying both sides of the equation (\ref{minRS3}) by a factor
$(n-1)\Omega_{n-1}\tilde{r}_{A}^{n-1}\left(1-\frac{\dot {\tilde
{r}_A}}{2H\tilde r_A}\right)$, and using the expression
(\ref{surgrav}) for the surface gravity, after some simplification
one can rewrite this equation in the form
\begin{eqnarray}
\label{minRS4}
-\frac{\kappa}{2\pi}\frac{(n-1)}{2G_{n+1}}\Omega_{n-1}\tilde
{r}_{A}^{n-2}d\tilde {r}_{A}&=&(n-1)\Omega_{n-1}\tilde
 {r}_{A}^{n-1}H(\rho+p)dt \nonumber \\
 &&-\frac{(n-1)}{2}\Omega_{n-1}\tilde
 {r}_{A}^{n-2}(\rho+p)d\tilde {r}_{A}.
 \end{eqnarray}
$E=\rho V$ is the total energy of the matter inside the
$(n-1)$-sphere of radius $\tilde{r}_{A}$ on the brane, where
$V=\Omega_{n-1}\tilde{r}_{A}^{n-1}$ is the volume enveloped by
$(n-1)$- dimensional sphere with the area of apparent horizon
$A=(n-1)\Omega_{n-1}\tilde{r}_{A}^{n-2}$  and
$\Omega_{n-1}=\frac{\pi^{(n-1)/2}}{\Gamma((n+1)/2)}$. Taking
differential form of the relation $ E=\rho
\Omega_{n-1}\tilde{r}_{A}^{n-1}$ for the total matter energy
inside the apparent horizon on the brane, we get
\begin{equation}
\label{dE1}
 dE=(n-1)\Omega_{n-1}\tilde
 {r}_{A}^{n-2}\rho d\tilde {r}_{A}+\Omega_{n-1}\tilde
 {r}_{A}^{n-1}\dot {\rho} dt.
\end{equation}
Using the continuity relation (\ref{Cont}) we obtain
\begin{equation}
\label{dE2}
 dE=(n-1)\Omega_{n-1}\tilde
 {r}_{A}^{n-2}\rho d\tilde {r}_{A}-(n-1)H(\rho+p)\Omega_{n-1}\tilde
 {r}_{A}^{n-1}dt.
\end{equation}
Substituting this relation into (\ref{minRS4}), and using the
relation between temperature and the surface gravity, we get the
first law of thermodynamics on the apparent horizon
\begin{equation}\label{FirminRS}
dE = TdS + WdV,
\end{equation}
where $W=(\rho-p)/2$ is the matter work density which is defined
by $W=-\frac{1}{2} t^{\mu \nu}h_{\mu \nu}$ \cite{Hay2}, and the
entropy of the apparent horizon on the brane is now given by
\begin{eqnarray}\label{entminRS}
S = {\displaystyle \int^{\tilde r_A}_0 dS }
&=&\frac{(n-1)\Omega_{n-1}}{2G_{n+1}}{\displaystyle \int^{\tilde
r_A}_0 \tilde {r}_{A}^{n-2}d\tilde
{r}_{A}}=\frac{2\Omega_{n-1}\tilde {r}_{A}^{n-1}}{4G_{n+1}}.
\end{eqnarray}
Let us note that the entropy obeys the area formula of horizon in
the bulk (the factor $2$ comes from the $\mathbb{Z}_2$ symmetry in
the bulk). This is due to the fact that because of the absence of
the negative cosmological constant in the bulk, no localization of
gravity happens on the brane. As a result, the gravity on the
brane is still $(n+1)$-dimensional.

On the other hand, from the global point of view, the apparent
horizon on the brane can be extended into the bulk. Since the
gravity in the bulk is described by pure $(n+1)$-dimensional
Einstein theory, then one can obtain the entropy of the apparent
horizon by the area formula in the bulk. We directly calculate the
area of the apparent horizon which extends into the bulk, and give
the entropy expression of apparent horizon from the bulk geometry.
Setting $\Lambda_{n+1}=\mathcal{C}=0$ in Eq. (\ref{Field}), this
equation reduces to
\begin{equation}\label{FieldMRS}
{A^{\prime}}^2-\alpha^2(t)-k=0,
\end{equation}
with the solution
\begin{equation}\label{AMRS}
A(t,y)=a(t)(1-\frac{|y|}{\tilde{r}_A})=a(t)f(\tilde{r}_A, y),
\end{equation}
where we have used $\alpha(t)=\dot{A}/N=\dot{a}$, since
$N(t,0)=1$. The function $f(\tilde{r}_A, y)$ has a positive root
at $y_0=\tilde{r}_A$. Then, the area of the apparent horizon in
the $(n+1)$-dimensional bulk (assuming $\mathbb{Z}_2$ symmetry )
can be written as (see \cite{Cai4} for details)
\begin{equation}
\mathcal{A}=2 \times
(n-1)\Omega_{n-1}\tilde{r}_A^{n-2}\int_0^{y_0}f^{n-2}(\tilde{r}_A,y)dy=2\Omega_{n-1}\tilde{r}_A^{n-1}.
\end{equation}
According to the $(n+1)$-dimensional area formula, we obtain the
entropy in the bulk
\begin{equation}
S=\frac{\mathcal{A}}{4 G_{n+1}}=\frac{2\Omega_{n-1}\tilde
{r}_{A}^{n-1}}{4G_{n+1}}.
\end{equation}
This expression for the entropy is exactly the same as the entropy
expression (\ref{entminRS}) which we have derived at the apparent
horizon on the brane from the first law of thermodynamics.

\subsection{Brane embedded in AdS bulk}\label{AdSRS}

In subsection \ref{minRS} we have assumed that the bulk cosmological
constant is absent. Here we leave that assumption, and further we
suppose that $\Lambda_{n+1}<0$. Using Eq. (\ref{rela}) the Friedmann
equation (\ref{RSFri}) can be written as
\begin{equation}\label{RSAd1}
 \sqrt{H^2+\frac{k}{a^2}+\frac{1}{\ell^2}}= \frac{4\pi
 G_{n+1}}{n-1}\rho.
 \end{equation}
In terms of the apparent horizon radius we have
\begin{equation}\label{RSAd2}
 \rho = \frac{n-1}{4\pi G_{n+1}}\sqrt{\frac{1}{{\tilde{r}_A}^2}+\frac{1}{\ell^2}},
\end{equation}
Taking differential form of the equation (\ref{RSAd2}) and using
the continuity equation (\ref{Cont}), one gets the differential
form of the Friedmann equation on the brane
\begin{equation}
\label{RSAd3} \label{RSAd3} H(\rho+p)dt=\frac{\ell}{4\pi
G_{n+1}{\tilde{r}_A}^2}\frac{d
\tilde{r}_A}{\sqrt{{\tilde{r}_A}^2+\ell^2}}.
\end{equation}
Again, multiplying both sides of equation (\ref{RSAd3}) by a
factor $(n-1)\Omega_{n-1}\tilde{r}_{A}^{n-1}\left(1-\frac{\dot
{\tilde r}_A}{2H\tilde r_A}\right)$, and using Eqs.
(\ref{surgrav}) and (\ref{dE2}), after some simplifications one
can rewrite this equation in the form
\begin{equation}\label{Fridif4}
dE -WdV
=\frac{\kappa}{2\pi}\frac{(n-1)\ell}{2G_{n+1}}\frac{\Omega_{n-1}\tilde
{r}_{A}^{n-2}}{\sqrt{{\tilde{r}_A}^2+\ell^2}}d\tilde {r}_{A}.
\end{equation}
This expression is nothing, but the first law of thermodynamics at
the apparent horizon on the brane, namely $dE=TdS+WdV$. We can
define the entropy associated with the apparent horizon on the
brane as
\begin{equation}
\label{entRSAdS1} \label{entRSAdS1} S=\frac{(n-1)\ell
\Omega_{n-1}}{2G_{n+1}}{\displaystyle\int^{\tilde
r_A}_0\frac{\tilde{r}_A^{n-2}
}{\sqrt{\tilde{r}_A^2+\ell^2}}d\tilde{r}_A}.
\end{equation}
Finally, the explicit form of the entropy at the apparent horizon
can be obtained by integrating (\ref{entRSAdS1}). The result is
\begin{equation} \label{entRSAdS2}
S=\frac{2\Omega_{n-1}{\tilde{r}_A}^{n-1}}{4 G_{n+1}}
 \times
{}_2F_1\left(\frac{n-1}{2},\frac{1}{2},\frac{n+1}{2},
-\frac{{\tilde{r}_A}^2}{\ell^2}\right),
\end{equation}
where ${}_2F_1(a,b,c,z)$ is a hypergeometric function. This
expression for the entropy is in complete agreement with the entropy
expression obtained in \cite{Cai4} by using the method of unified
first law and Clausius relation. We stress here that, in order to
obtain an entropy expression on the brane one does not need to use
the unified first law and Clausius relation. One can directly write
down the Friedmann equation in the form of first law and extract the
entropy expression.  Also, it is easy to check that if one writes
the Friedmann equation in RS II model in the form of \cite{Cai4}
($\rho= \rho_{b}+\lambda$), then following the above method one can
rewrite the Friedmann equation directly in the form of the first
law, with entropy expression given by (\ref{entRSAdS2}). It is worth
noticing when $\tilde{r}_A \ll\ell$, which physically means that the
size of the extra dimension is very large if compared with the
apparent horizon radius, one recovers the $(n+1)$-dimensional area
formula for the entropy on the brane $S
=2\Omega_{n-1}{\tilde{r}_A}^{n-1}/4G_{n+1}$. The factor $2$ comes
from the $\mathbb{Z}_2$ symmetry in the bulk. This is an expected
result since in this regime we have a quasi- Minkowski bulk and we
have shown in the previous subsection that for a RS II brane
embedded in the Minkowski bulk, the entropy on the brane follows the
$(n+1)$-dimensional area formula in the bulk. One can also extend
the apparent horizon on the brane into the bulk and determine the
apparent horizon entropy by using the $(n+1)$-dimensional area
formula in the bulk. To do this, we put $\mathcal{C}=0$ in Eq.
(\ref{Field}) and use Eq. (\ref{rela}), so
\begin{equation}\label{FieldAdSRS}
{A^{\prime}}^2-\frac{A^2}{\ell^2}-\alpha^2(t)-k=0.
\end{equation}
This equation has a solution of the form (see \cite{BDEL} for
details)
\begin{equation}\label{AAdSRS}
A(t,y)=a(t)\left(-\frac{1}{2}\frac{\ell^2}{\tilde{r}_A^2}
+\left(1+\frac{1}{2}\frac{\ell^2}{\tilde{r}_A^2}\right)\cosh{\left(\frac{2y}{\ell}\right)}
-\sqrt{1+\frac{\ell^2}{\tilde{r}_A^2}}\sinh{\left(\frac{2|y|}{\ell}\right)}\right)^{\frac{1}{2}}.
\end{equation}
The function $f(\tilde{r}_A, y)$ has a positive root at $y_0=\ell
\mathrm{sinh}^{-1}(\frac{\tilde{r}_A}{\ell})$. Then, the area of
the apparent horizon in the $(n+1)$-dimensional bulk (assuming
$\mathbb{Z}_2$ symmetry) can be written as
\begin{eqnarray}
\mathcal{A}&=&2 \times
(n-1)\Omega_{n-1}\tilde{r}_A^{n-2}\int_0^{y_0}f^{n-2}(\tilde{r}_A,y)dy
\nonumber\\
 &=&2 \times \Omega_{n-1}{\tilde{r}_A}^{n-1}
 \times {}_2F_1\left(\frac{n-1}{2},\frac{1}{2},\frac{n+1}{2},
-\frac{{\tilde{r}_A}^2}{\ell^2}\right),
\end{eqnarray}
According to the $(n+1)$-dimensional area formula, we obtain the
entropy in the bulk
\begin{eqnarray}\label{entbulkAdS}
S&=&\frac{\mathcal{A}}{4 G_{n+1}}=\frac{2\Omega_{n-1}\tilde
{r}_{A}^{n-1}}{4G_{n+1}}\times
{}_2F_1\left(\frac{n-1}{2},\frac{1}{2},\frac{n+1}{2},
-\frac{{\tilde{r}_A}^2}{\ell^2}\right).
\end{eqnarray}
This expression for the entropy is exactly the same as the entropy
expression (\ref{entRSAdS2}) which we have derived at the apparent
horizon on the brane from the first law of thermodynamics. Indeed,
we have shown that for RS II brane world embedded in
$(n+1)$-dimensional (Minkowski) AdS bulk, the entropy in the bulk
is exactly the same as the entropy expression associated with
apparent horizon on the brane. This is in agreement with the
arguments in \cite{Emp}.

\section{Thermodynamics of Apparent Horizon in DGP model}\label{DGP}

In the previous section, we have studied thermodynamical behavior of
Friedmann equation at apparent horizon for RS II brane world
embedded in $(n+1)$-dimensional Minkowski and AdS bulks and have
showed that the Friedmann equation can be written directly in the
form of first law. We have found an explicit expression for the
entropy of the apparent horizon on the brane. In this section, we
are going to extend the discussion to the case in which the
intrinsic curvature term of the brane is included in the action,
namely DGP brane world. The generalized Friedmann equation for DGP
model is given in Eq. (\ref{GFri}). Again, we are interested in
studying DGP brane world embedded in the Minkowski and AdS bulks, so
$\mathcal{C}=0$.

\subsection{Brane embedded in Minkowski bulk}\label{minDGP}

In the Minkowski bulk, $\Lambda_{n+1}=0$, and the Friedmann equation
(\ref{GFri}) reduces to the form
\begin{equation}\label{DGmin1}
\sqrt{H^2+\frac{k}{a^2}}
=-\frac{\kappa_{n+1}^2}{4\kappa_{n}^2}(n-2)(H^2+\frac{k}{a^2})+\frac{\kappa_{n+1}^2}{2(n-1)}\rho.
\end{equation}
In terms of the apparent horizon radius, we can rewrite this
equation in the form
\begin{equation}\label{DGmin2}
 \rho = \frac{(n-1)(n-2)}{2\kappa_{n}^2}\frac{1}{{\tilde{r}_A}^2}+\frac{2(n-1)}{\kappa_{n+1}^2}\frac{1}{{\tilde{r}_A}}
 .\end{equation}
Taking the differential form of the equation (\ref{DGmin2}) and
using the continuity equation (\ref{Cont}), one gets the
differential form of the Friedmann equation on the brane
\begin{equation}\label{DGmin3}
 H(\rho+p)dt = \frac{n-2}{8\pi G_n}\frac{d \tilde{r}_A}{{\tilde{r}_A}^3}+\frac{1}{4 \pi G_{n+1}}\frac{d\tilde{r}_A}{{\tilde{r}_A}^2},
\end{equation}
where we have used Eq. (\ref{rela}). Now, we multiply both sides
of the equation (\ref{DGmin3}) by a factor
$(n-1)\Omega_{n-1}\tilde{r}_{A}^{n-1}\left(1-\frac{\dot {\tilde
r}_A}{2H\tilde r_A}\right)$, and use Eqs. (\ref{surgrav}) and
(\ref{dE2}), then we can rewrite this equation in the form of the
first law
\begin{equation}\label{DGmin4}
dE - WdV = \frac{\kappa}{2\pi}
(n-1)\Omega_{n-1}\left(\frac{(n-2){\tilde{r}_A}^{n-3}}{4G_n}+\frac{{\tilde{r}_A}^{n-2}}{2G_{n+1}}
\right)d\tilde{r}_A=TdS,
\end{equation}
where the entropy can be given by
\begin{eqnarray}\label{entDGmin1}
S &=& (n-1)\Omega_{n-1}{\displaystyle\int^{\tilde
r_A}_0\left(\frac{(n-2){\tilde{r}_A}^{n-3}}{4G_n}+\frac{{\tilde{r}_A}^{n-2}}{2G_{n+1}}
\right)d\tilde{r}_A} \nonumber \\
&=&\frac{(n-1)\Omega_{n-1}{\tilde{r}_A}^{n-2}}{4G_n}+\frac{2\Omega_{n-1}{\tilde{r}_A}^{n-1}}
{4G_{n+1}}=S_{n}+S_{n+1}.
\end{eqnarray}
It is interesting to note that in this case the entropy can be
regarded as a sum of two area formulas; one (the first term)
corresponds to the gravity on the brane and  the other (the second
term) to the gravity in the bulk. This indeed reflects the fact
that there are two gravity terms in the action of DGP model.

\subsection{Brane embedded in AdS bulk}\label{AdSDGP}

For the AdS bulk, $\Lambda_{n+1}<0$, we can write the Friedmann
equation (\ref{GFri}) in the form
\begin{equation}\label{DGAdS1}
\sqrt{H^2+\frac{k}{a^2}+\frac{1}{\ell^2}}
=-\frac{\kappa_{n+1}^2}{4\kappa_{n}^2}(n-2)(H^2+\frac{k}{a^2})+\frac{\kappa_{n+1}^2}{2(n-1)}\rho,
\end{equation}
where we have used Eq. (\ref{rela}). In terms of the apparent
horizon radius, this equation can be rewritten as
\begin{equation}\label{DGAdS2}
 \rho = \frac{(n-1)(n-2)}{2\kappa_{n}^2}\frac{1}{{\tilde{r}_A}^2}+\frac{2(n-1)}{\kappa_{n+1}^2}
 \sqrt{\frac{1}{{\tilde{r}_A}^2}+\frac{1}{\ell^2}}.
\end{equation}
If one takes the differential form of the equation (\ref{DGAdS2}),
after using Eqs. (\ref{rela}) and (\ref{Cont}), one gets the
differential form of the Friedmann equation on the brane
 \begin{equation}\label{DGAdS3}
 H(\rho+p)dt = \frac{n-2}{8\pi G_n}\frac{d \tilde{r}_A}{{\tilde{r}_A}^3}+\frac{\ell}{4\pi G_{n+1}}\frac{d\tilde{r}_A}{{\tilde{r}_A}^2}
 \frac{1}{\sqrt{{\tilde{r}_A}^2+\ell^2}} ,
\end{equation}
Again, multiplying both sides of the equation (\ref{DGAdS3}) by a
factor $(n-1)\Omega_{n-1}\tilde{r}_{A}^{n-1}\left(1-\frac{\dot
{\tilde r}_A}{2H\tilde r_A}\right)$, and using Eqs.
(\ref{surgrav}) and (\ref{dE2}), one gets
\begin{equation}
 dE - WdV = \frac{\kappa}{2\pi}
(n-1)\Omega_{n-1}\left(\frac{(n-2){\tilde{r}_A}^{n-3}}{4G_n}+\frac{\ell}{2G_{n+1}}\frac{{\tilde{r}_A}^{n-2}}{\sqrt{{\tilde{r}_A}^2+\ell^2}}
\right)d\tilde{r}_A.
\end{equation}
One can immediately see that this equation has the form of the
first law $dE=TdS+WdV$, if one writes entropy associated with the
apparent horizon on the brane as
\begin{eqnarray}\label{entDGAdS1}
 S &=&
(n-1)\Omega_{n-1}{\displaystyle\int^{\tilde
r_A}_0\left(\frac{(n-2){\tilde{r}_A}^{n-3}}{4G_n}+\frac{\ell}{2G_{n+1}}\frac{{\tilde{r}_A}^{n-2}}{\sqrt{{\tilde{r}_A}^2+\ell^2}}
\right)d\tilde{r}_A}.
\end{eqnarray}
Integrating this equation one gets the explicit form for the
entropy at the apparent horizon
\begin{equation} \label{entRSAdS}
S=\frac{(n-1)\Omega_{n-1}{\tilde{r}_A}^{n-2}}{4G_{n}}+\frac{2\Omega_{n-1}{\tilde{r}_A}^{n-1}}{4G_{n+1}
}
 \times
{}_2F_1\left(\frac{n-1}{2},\frac{1}{2},\frac{n+1}{2},
-\frac{{\tilde{r}_A}^2}{\ell^2}\right).
\end{equation}
Again, we see that in the warped DGP brane model embedded in the AdS
bulk, the entropy associated with the apparent horizon on the brane
has two parts, $S=S_{n}+S_{n+1}$. The first part $S_{n}$, which
follows the $n$-dimensional area law on the brane and the second
part $S_{n+1}$ which is the same as the entropy expression obtained
by extension the apparent horizon into the bulk (see Eq.~(
\ref{entbulkAdS})) and therefore obeys the $(n+1)$-dimensional area
law in the bulk.

\section{ Conclusion}\label{sum}

To summarize, we have showed that the Friedmann equations on the
$(n-1)$-dimensional brane embedded in an $(n+1)$-dimensional
spacetime can be written directly in the form of the first law of
thermodynamics, $dE=TdS+WdV$. Note that here $E$ is not the
Misner-Sharp energy, but the matter energy $\rho V$ inside the
apparent horizon  and they are equal only in Einstein
gravity~\cite{Cai2,Cai3,Cai4}. This procedure leads to extract an
expression for the entropy at the apparent horizon on the brane,
which is useful in studying the thermodynamical properties of the
black hole horizon on the brane. We have discussed several cases
including whether there is or not a cosmological constant in the
bulk and whether there is or not an intrinsic curvature term on the
brane. Interestingly enough, we have noted that when the
cosmological constant vanishes in the bulk and the intrinsic
curvature term is absent, the entropy of apparent horizon on the
brane obeys the area formula (\ref{entminRS}) in the bulk. This is
actually expected since the brane looks like a domain wall moving in
a Minkowski spacetime; no localization of gravity happens in this
case, unlike the case of RS II model, where the localization of
gravity occurs due to the negative cosmological constant in the
bulk. Another interesting point we found in this paper is that when
the intrinsic curvature of the brane is added to the action, the
entropy expression of the apparent horizon will include a term which
satisfies the area formula on the brane. This can also be understood
because the Einstein-Hilbert term on the brane contributes the area
term.

\acknowledgments This work was partially supported by NNSF of
China, Ministry of Education of China and Shanghai Educational
Commission. RGC was supported partially by grants from NSFC, China
(No. 10325525 and No. 90403029), and a grant from the Chinese
Academy of Sciences. RGC thanks C.M. Cao for helpful discussions.

\end{document}